\documentclass[
reprint,            % journal's actual layout
superscriptaddress, % authors with affiliations via superscripts
amsmath,            % add AMS-Latex features
amssymb,            % add extra AMS symbols, including amsfonts
aps,                % aps or aip
prd,                 % prl, pra, prb, prc, prd, pre, prstab
floatfix,           % process floats as early as possible
nofootinbib,        % show footnote at the text
]{revtex4-1}

\usepackage{tensor}     % manipulate tensors
\usepackage{graphicx}   % include figures
\usepackage[
colorlinks=true,        % color link
citecolor=blue,         % cite color
linkcolor=blue,         % link color
urlcolor=blue           % url color
]{hyperref}             % create hyperlinks
\usepackage{bm}         % \bm{<text>} Bold math symbols
\usepackage{xcolor}     % textcolor
\def\({\left(}
\def\){\right)}
\def\[{\left[}
\def\]{\right]}

\def\e{\begin{equation}}
\def\q{\end{equation}}
\def\m{\begin{eqnarray}}
\def\n{\end{eqnarray}}

\def\a{\begin{aligned}}
\def\b{\end{aligned}}

\newcommand{\nc}{\newcommand*}
\nc{\Eq}[1]{Eq.~\eqref{#1}}     % equation
\nc{\Fig}[1]{Fig.~\ref{#1}}     % figure
\nc{\Table}[1]{Table~\ref{#1}}  % table
\nc{\Sec}[1]{Sec.~\ref{#1}}     % section
\nc{\Msun}{M_\odot}

\usepackage{color}

\begin{document}
\title{Gravitational and electromagnetic radiation from binary black holes with electric and magnetic charges: Elliptical orbits on a cone}
%%%%%%%%%%%%%%%%%%%%%%%%%%%%%%% author 1 %%%%%%%%%%%%%%%%%%%%%%%%%%%%%
\author{Lang Liu}
\email{liulang@itp.ac.cn}
\affiliation{CAS Key Laboratory of Theoretical Physics,
Institute of Theoretical Physics, Chinese Academy of Sciences,
Beijing 100190, China}
\affiliation{School of Physical Sciences,
University of Chinese Academy of Sciences,
No. 19A Yuquan Road, Beijing 100049, China}
%%%%%%%%%%%%%%%%%%%%%%%%%%%%%%% author 2 %%%%%%%%%%%%%%%%%%%%%%%%%%%%%
\author{{\O}yvind Christiansen}
\email{oyvind.christiansen@astro.uio.no}
\affiliation{Institute of Theoretical Astrophysics, University of Oslo, Sem S$\ae$lands vei 13,0371 Oslo, Norway}

%%%%%%%%%%%%%%%%%%%%%%%%%%%%%%% author 3 %%%%%%%%%%%%%%%%%%%%%%%%%%%%%

\author{Wen-Hong Ruan}
\email{ruanwenhong@itp.ac.cn}
\affiliation{CAS Key Laboratory of Theoretical Physics,
Institute of Theoretical Physics, Chinese Academy of Sciences,
Beijing 100190, China}
\affiliation{School of Physical Sciences,
University of Chinese Academy of Sciences,
No. 19A Yuquan Road, Beijing 100049, China}

%%%%%%%%%%%%%%%%%%%%%%%%%%%%%%% author 4 %%%%%%%%%%%%%%%%%%%%%%%%%%%%%
\author{Zong-Kuan Guo}
\email{guozk@itp.ac.cn}
\affiliation{CAS Key Laboratory of Theoretical Physics,
Institute of Theoretical Physics, Chinese Academy of Sciences,
Beijing 100190, China}
\affiliation{School of Physical Sciences,
University of Chinese Academy of Sciences,
No. 19A Yuquan Road, Beijing 100049, China}
\affiliation{School of Fundamental Physics and Mathematical Sciences,
Hangzhou Institute for Advanced Study,
University of Chinese Academy of Sciences, Hangzhou 310024, China}

%%%%%%%%%%%%%%%%%%%%%%%%%%%%%%% author 5 %%%%%%%%%%%%%%%%%%%%%%%%%%%%%
\author{Rong-Gen Cai}
\email{cairg@itp.ac.cn}
\affiliation{CAS Key Laboratory of Theoretical Physics,
Institute of Theoretical Physics, Chinese Academy of Sciences,
Beijing 100190, China}
\affiliation{School of Physical Sciences,
University of Chinese Academy of Sciences,
No. 19A Yuquan Road, Beijing 100049, China}
\affiliation{School of Fundamental Physics and Mathematical Sciences,
Hangzhou Institute for Advanced Study,
University of Chinese Academy of Sciences, Hangzhou 310024, China}
%%%%%%%%%%%%%%%%%%%%%%%%%%%%%%% author 6 %%%%%%%%%%%%%%%%%%%%%%%%%%%%%
\author{Sang Pyo Kim}
\email{sangkim@kunsan.ac.kr}
\affiliation{CAS Key Laboratory of Theoretical Physics,
Institute of Theoretical Physics, Chinese Academy of Sciences,
Beijing 100190, China}
\affiliation{Department of Physics, Kunsan National University, Kunsan 54150, Korea}
\date{\today}
%%%%%%%%%%%%%%%%%%%%%%%%%%%%%%%%%%%%%%%%%%%%%%%%%%%%%%%%%%%%%%%%%%%%%%
\begin{abstract}
%By using a Newtonian method with radiation reactions, we calculate the total emission rate of energy and angular momentum due to gravitational and electromagnetic radiation from binary black holes with electric and magnetic charges for precessing elliptical orbits. It is shown that the emission rates of energy and angular momentum due to gravitational radiation and electromagnetic radiation have the same dependence on the conic angle for different orbits. Moreover, we obtain the evolutions of orbits and find that a circular orbit remains circular and an elliptic orbit becomes quasi-circular due to electromagnetic and gravitational radiation. Using the evolution of orbits, we derive the waveform models for dyonic binary black hole inspirals which can be used to investigate whether black holes have electric and magnetic charges or not.

 Extending the electromagnetic and gravitational radiations from binary black holes with electric and magnetic charges in circular orbits in Phys. Rev. D {\bf 102}, 103520 (2020), we calculate the total emission rates of energy and angular momentum due to gravitational and electromagnetic radiations from dyonic binary black holes in precessing elliptical orbits. It is shown that the emission rates of energy and angular momentum due to gravitational and electromagnetic radiations have the same dependence on the conic angle for different orbits. Moreover, we obtain the evolutions of orbits and find that a circular orbit remains circular while an elliptic orbit becomes quasi-circular due to electromagnetic and gravitational radiations. Using the evolution of orbits, we derive the waveform models for dyonic binary black hole inspirals and show the amplitudes of the gravitational waves for dyonic binary black hole inspirals differ from those for Schwarzschild binary black hole inspirals, which can be used to test electric and magnetic charges of black holes.
\end{abstract}

\maketitle
%%%%%%%%%%%%%%%%%%%%%%%%%%%%%%%%%%%%%%%%%%%%%%%%%%%%%%%%%%%%%%%%%%%%%%
%%%%%%%%%%%%%%%%%%%%%%%%%%%%%%%%%%%%%%%%%%%%%%%%%%%%%%%%%%%%%%%%%%%%%%
\section{Introduction}\label{intro}
Magnetic charges, if they exist in the early Universe, will provide a hitherto unexplored window to probe fundamental physics in the Standard Model of particle physics and beyond.
Though no evidence of magnetic charges has been found yet ~\cite{Staelens:2019gzt,2020arXiv200505100M}, magnetically charged black holes have attracted much attention not only in theoretical study but also in recent astronomical observations \cite{Maldacena:2020skw,Bai:2020spd,Ghosh:2020tdu}. Recently, spectacular properties of magnetic charged black holes have been extensively discussed in \cite{Maldacena:2020skw}. It is shown there that the magnetic field near the horizon of a magnetic black hole could be strong enough to restore the electroweak symmetry. The phenomenology of low-mass magnetic black holes, which can have electroweak-symmetric coronas outside of the event horizon, has been comprehensively studied in \cite{Bai:2020spd}. Potential astrophysical signatures for magnetically charged black holes have also been investigated in \cite{Ghosh:2020tdu}.

According to the ``no-hair" conjecture, a general relativistic black hole is completely described by four parameters: mass, angular momentum, magnetic charge as well as electric charge. Compared to Schwarzschild black holes, charged black holes have rich phenomena. Recently, charged black holes have been discussed extensively \cite{Cardoso:2016olt,Liebling:2016orx,Toshmatov:2018tyo,Bai:2019zcd,Allahyari:2019jqz,Liu:2020cds,Christiansen:2020pnv,Wang:2020fra,Bozzola:2020mjx,Kim:2020bhg,Cardoso:2020nst,Christiansen:2020wzi,Cardoso:2020iji, McInnes:2020gxx}. Binary black holes with charges emit not only gravitational radiation but also electromagnetic radiation.
By using a Newtonian method with the inclusion of radiation reaction, a previous study \cite{Liu:2020cds} obtains the evolutions of orbits and calculates merger times of binary black holes with electric charges by considering the Keplerian motion of two charged bodies and accounting for the loss of energy and angular momentum due to the emission of gravitational and electromagnetic waves. {The Coulomb-type force due to a pure electric or magnetic charge changes the coupling parameter of the gravitational force} and alters the gravitational wave emission compared to an uncharged binary, which leads to a different merger rate of primordial black hole \cite{Liu:2020cds}. The bias in the binary parameters due to the charge-chirp mass degeneracy is discussed in \cite{Christiansen:2020pnv}.
For the first merger event of binary black holes reported by LIGO/Virgo, GW150914, Ref.~\cite{Liebling:2016orx} argues that the magnetic charge should be small, and it is shown in Ref.~\cite{Wang:2020fra} that binary black holes could have some electric charge by using a full Bayesian analysis with Gaussian noise.

A dyonic black hole is a {nonrotating or rotating black hole with an electric charge $q$ and a magnetic charge $g$. A dyonic nonrotating black hole} has the same metric as the Reissner-Nordstr\"{o}m black hole with $q^2$ replaced by $q^2+g^2$ \cite{1982PhRvD..25..995K}.
In the Minkowski spacetime, the nonrelativistic interaction of two dyons  was studied in a quantum theory~\cite{1968PhRv..176.1480Z} and in classical theory~\cite{1976AnPhy.101..451S} .
 {In the previous paper \cite{Liu:2020vsy}, which will be denoted as I, we have derived the equations of motion of dyonic black hole binaries and explored features of static orbits (without radiation)}. In Ref.~I, static orbits of dyonic black hole binaries {on a conic section are divided into three categories: (1) $e=0$; (2) $e \neq 0$ and $\sin \theta$ is rational; (3) $e \neq 0$ and $\sin \theta$ is irrational ($\theta$ is the conic angle, $e$ is the eccentricity with the definition in Eq.~\eqref{tilde}.)}, and the orbits of dyonic black hole binaries in those different cases have different topology. In the first case of $e=0$, the three-dimensional trajectory is a two-dimensional circular orbit. In the second case when $e \neq 0$ and $\sin \theta$ is rational, the orbit is closed and confined to the surface of a cone. In the last case when $e \neq 0$ and $\sin \theta$ is irrational, the conic-shaped orbit of the binary is not closed and shows a chaotic behavior of a conserved autonomous system. {For circular orbits, in Ref. I we have calculated the total emission rate of energy and angular momentum due to gravitational and electromagnetic radiation. Furthermore, the merger times of dyonic binaries for circular orbits are calculated.}

In the universe, most binary black hole systems have non-zero eccentricity and can have even large eccentricity for those from encounters of black holes. Indeed the circular orbits are a very rare and special case. Therefore, it is important and meaningful to explore the evolutions of elliptical orbits for binary black holes with electric and magnetic charges and their characteristic features.
In this paper, we explore the evolutions of elliptical orbits for binary black holes with electric and magnetic charges by considering the equation of motion of two dyonic bodies and accounting for a loss of energy via quadrupolar emission of gravitational waves and dipolar emission of electromagnetic ones.  For dyonic binaries, in the 0th order post-Newtonian (PN) approximation, the angular-momentum-dependent and non-central Lorentz force cause the orbits to execute three-dimensional and complex trajectories. We find that the emission rates of energy and angular momentum have the same dependence of $\theta$ for all different cases. Moreover, we find a circular orbit remains circular and an elliptic orbit becomes quasi-circular because of electromagnetic and gravitational radiation.  Using the evolution of orbits, we derive the waveform models for dyonic binary black hole inspirals.

The organization of this paper as follows. In \Sec{Emission}, we work out the total emission rate of angular momentum and energy due to gravitational and electromagnetic radiation. In \Sec{Evolution}, we derive the evolutions of orbits and find a circular orbit remains circular and an elliptic orbit becomes quasi-circular.  In \Sec{Waveformcompare}, we obtain the waveforms for dyonic binary black hole inspirals.
In \Sec{Concl}, we discuss the physical implications and conclude the perspective of the dyonic black hole binaries. Throughout this work, we set $G=c =4 \pi \varepsilon_{0} = \frac{\mu_0}{4\pi}=1$.

\section{Electromagnetic and Gravitational Radiation}
\label{Emission}

In this paper, we adopt the the weak-field approximation. In other words, we only focus on the case that the distance of the dyonic black hole binary is much larger than their event horizons. In such a case, a dyonic black hole binary during the inspiral motion whose distance is much larger than their event horizons is well approximated by a pair of massive point-like objects with electric and magnetic charges.

 The elliptical orbits on a cone which precess around the generalized angular momentum are a characteristic feature of the dyonic binary with both electric and magnetic charges. We extend the previous works~\cite{Liu:2020vsy} to inspiral elliptical orbits on a cone, find the gravitational radiation \'{a} la~\cite{Peters:1963ux,Peters:1964zz} and the synchrotron radiation due to electric and magnetic dipoles \'{a} la~\cite{1975ctf..book.....L} and investigate the radiation reaction on the orbital motion.

\subsection{$\sin \theta$ is rational}
\label{Rational}
In this subsection, we calculate the emissions of energy and angular momentum due to gravitational and electromagnetic radiation for the case $e \neq 0$ and $\sin \theta$ is rational. {To do so}, we set $
\sin \theta=\frac{l}{n}$, where $n$ and $l$ are relatively positive prime numbers and $l<n$.

Following \cite{Liu:2020vsy}, without radiation, when we choose the $z$-axis along $\bm{L}$, the orbit is given by \footnote{Throughout this work, we only consider $0<\theta\leq \pi/2$. For $\pi/2\leq\theta< \pi$, we can refine $\boldsymbol{R}^{\prime}=-\boldsymbol{R}$ to make $0<\theta\leq \pi/2$.}
\m
\label{R}
\boldsymbol{R}&=&\frac{\frac{\tilde{L}^{2}}{\mu|C|}}{1+\sqrt{1+\frac{2 \tilde{L}^{2}}{\mu C^{2}} E} \cos (\phi \sin \theta)} \left(\begin{array}{c}
\sin \theta \cos \phi \\
\sin \theta \sin \phi \\
\cos \theta
\end{array}\right)
\notag \\
&\equiv& \frac{a\left(1-e^{2}\right)}{1+e \cos (\phi \sin \theta)}\left(\begin{array}{c}
\sin \theta \cos \phi \\
\sin \theta \sin \phi \\
\cos \theta
\end{array}\right) ,
\n
where $a$ and $e$ {that can be interpreted as the semimajor axis and eccentricity} are defined by
\m
\label{E}
a \equiv \frac{C}{2E},
\n
\m
\label{tilde}
e \equiv \left(1-\frac{2E \tilde{L}^{2}}{\mu C^2}\right)^{1/2},
\n
and $C=\left(-\mu M+q_{1} q_{2}+g_{1} g_{2}\right),  D=\left(q_{2} g_{1}-g_{2} q_{1}\right)$. Here, $E$ is the orbital energy of the binary. For our bound system, $E<0$ which means $C<0$. According to \cite{shnir_magnetic_2005}, the generalized angular momentum of binary system $\bm{L}$ defined by
$\bm{L}\equiv \bm{\tilde{L}}-D \hat{\bm{r}}$,
 where $\hat{\bm{r}}$ is the unit vector along $\bm{R}$ and $\bm{\tilde{L}}\equiv \mu \bm{R} \times \bm{v}$ is the {orbital} angular momentum of binary system.

At first, we calculate the emission of electromagnetic radiation due to the electric and magnetic charges on the orbit \eqref{R}, averaged over one orbit.
Following \cite{Liu:2020vsy}, the energy emission due to electromagnetic radiation is given by
 \m \label{P-EM}
 P_{EM}=\frac{2 \mu^{2}((\Delta \sigma_q)^{2}+(\Delta \sigma_g)^{2})}{3} \ddot{R}^i \ddot{R}_i,
 \n
where $\mu=\frac{m_1m_2}{m1+m2}$ is the reduced mass and
\m
\Delta \sigma_q=q_{2} / m_{2}-q_{1} / m_{1},
\n
\m
\Delta \sigma_g=g_{2} / m_{2}-g_{1} / m_{1},
\n
are the dipole moments of electric charges and magnetic charges.
The averaged energy loss over an orbital period $T_2=\int_0 ^{2n\pi} d\phi \dot{\phi}^{-1}=2 \pi a^{3 / 2} \sqrt{-\mu / C} l$ due to  electromagnetic radiation is given by
\m
\label{A1}
&&\left\langle\frac{dE_{EM}}{dt}\right\rangle=-\bar{P}_{EM}=-\frac{1}{T_2} \int_0 ^{2n\pi} d\phi P_{EM} \dot{\phi}^{-1}
\notag \\
&=& -\frac{C^2((\Delta \sigma_q)^{2}+(\Delta \sigma_g)^{2}) n^2}{24 a^4 \left(1-e^2\right)^{5/2} l^2}
\notag \\
&\times& \left(3 e^4+\left(3 e^2+20\right) e^2 (1-\frac{2 l^2}{n^2})+28 e^2+16\right).
\n

The angular momentum emission due to electromagnetic radiation is given by
\m \label{L-EM}
\dot{L}_{EM}^{i}=-\frac{2 \mu^{2}((\Delta \sigma_q)^{2}+(\Delta \sigma_g)^{2})}{3} \epsilon_{j k}^{i}\dot{R}^{j} \ddot{R}^{k}.
\n
{Thus the angular momentum loss due to electromagnetic radiation averaged one orbital period $T_2$ is}
\m
\left\langle\frac{dL^i_{EM}}{dt}\right\rangle \equiv \frac{1}{T_2} \int_{0}^{T_2} d t \dot{L}^i_{EM}.
\n
After a straightforward computation, we obtain
\m
\left\langle\dot{L}_{EM}^1\right\rangle=\left\langle\dot{L}_{EM}^2\right\rangle=0,
\n
\m
\label{A2}
\left\langle\dot{L}_{EM}^3\right\rangle&=&-\frac{  (-C)^{3/2} \sqrt{\mu} ((\Delta \sigma_q)^{2}+(\Delta \sigma_g)^{2}) n }{6 a^{5/2} (1-e^2) l}
 \notag \\
&\times& \Bigl( e^2 (2-\frac{2 l^2}{n^2} )+4 \Bigr).
\n

Now, we compute the total radiated power in gravitational waves.  In our reference frame where $\bm{L}$ along the $z$ axis, the second mass moment is written as
\m
\label{Mij}
M^{ij}=\mu R^i R^j.
\n
Following \cite{Peters:1963ux}, the radiated power of gravitational waves can be expressed as
\m
P_{GW}=\frac{1}{5 }\left\langle\ddot{M}_{i j} \ddot{M}_{i j}-\frac{1}{3}\left(\ddot{M}_{k k}\right)^{2}\right\rangle
\n
A well-defined quantity of energy of gravitational waves is the average of $P_{GW}$ over one period $T_2$
\m
\bar{P}_{GW} \equiv \frac{1}{T_2} \int_{0}^{T_2} d t P_{GW}.
\n
Thus the averaged energy loss over an orbital period $T_2$ is given by
\m
\label{A3}
&&\left\langle\frac{dE_{GW}}{dt}\right\rangle=-\bar{P}_{GW}=\frac{(-C)^3 n^4}{240 a^5 \left(1-e^2\right)^{7/2} l^4 \mu }
\notag \\
&\times& \Bigl( 2 \left(e^2+1\right) \left(15 e^2+308\right) e^2 (\frac{8 l^4}{n^4}-\frac{8 l^2}{n^2}+1)
\notag \\
&+&\left(-15 e^6+26 e^4+1976 e^2+720\right) (1-\frac{2 l^2}{n^2})
\notag \\
&-&3 \left(15 e^6+404 e^4+1104 e^2+272\right) \Bigr).
\n

Following \cite{Peters:1964zz}, the rate of angular momentum emission due to gravitational waves is given by
\m
\frac{d L^{i}_{GW}}{d t}=-\frac{2}{5} \epsilon^{i k l}\left\langle\ddot{M}_{k a} \ddot{M}_{l a}\right\rangle
\n
We obtain the angular momentum loss due to gravitational radiation averaged one orbital period $T_2$
\m
\left\langle\frac{dL^i_{GW}}{dt}\right\rangle \equiv \frac{1}{T_2} \int_{0}^{T_2} d t \dot{L}^i_{GW}.
\n
It is straightforward to show
\m
\left\langle\dot{L}_{GW}^1\right\rangle=\left\langle\dot{L}_{GW}^2\right\rangle=0,
\n
and
\m
\label{A4}
&&\left\langle\dot{L}_{GW}^3\right\rangle=\frac{(-C)^{5/2 } n^3}{40 a^{7/2} \sqrt{\mu }(1-e^2)^2 l^3}
\notag \\
&\times& \Bigl( \left(7 e^2+48\right) e^2 (\frac{8 l^4}{n^4}-\frac{8 l^2}{n^2}+1)-2 \left(5 e^4+92 e^2+68\right)
\notag \\
&+&\left(-3 e^4+88 e^2+120\right) (1-\frac{2 l^2}{n^2}) \Bigr).
\n

\subsection{$\sin \theta$ is irrational}
\label{Irrational}
As shown in \cite{Liu:2020vsy}, when $e \neq 0$ and $\sin \theta$ is irrational, the orbit is not closed. In this subsection, we calculate the emissions of energy and angular momentum due to electromagnetic and gravitational radiation for the case of $e \neq 0$ and a irrational $\sin \theta$.

The energy emission (\ref{P-EM}) due to electromagnetic radiation leads to the averaged energy loss rate
\m
\bar{P}_{EM}=\lim _{ N\rightarrow \infty }\frac{1}{N T_3} \int_0 ^{2N\pi/\sin \theta} d\phi P_{EM} \dot{\phi}^{-1},
\n
where $ T_3=\int_0 ^{2\pi/\sin(\theta)} d\phi \dot{\phi}^{-1}=2 \pi a^{3 / 2} \sqrt{-\mu / C}.$
So, we can get
\m
\label{B1}
&&\left\langle\frac{dE_{EM}}{dt}\right\rangle=-\bar{P}_{EM}=- \frac{C^2((\Delta \sigma_q)^{2}+(\Delta \sigma_g)^{2}) \csc ^2(\theta )}{24 a^4 \left(1-e^2\right)^{5/2} }
 \notag \\
 &\times& \Bigl(3 e^4+\left(3 e^2+20\right) e^2 \cos (2\theta)+28 e^2+16 \Bigr).
\n
Similarly, the angular momentum emission (\ref{L-EM}) due to electromagnetic radiation
gives the angular momentum loss rate due to electromagnetic radiation
\m
\left\langle\dot{L}^{i}_{EM}\right\rangle=\lim _{ N\rightarrow \infty }\frac{1}{N T_3} \int_0 ^{2N\pi/\sin \theta} d\phi \dot{L}^{i}_{EM} \dot{\phi}^{-1}.
\n
After a straightforward computation, we have
\m
\label{B2}
\left\langle\dot{L}_{EM}^3\right\rangle&=&-\frac{  (-C)^{3/2} \sqrt{\mu} ((\Delta \sigma_q)^{2}+(\Delta \sigma_g)^{2}) \csc (\theta ) }{6 a^{5/2} (1-e^2) }
 \notag \\
&\times& \(e^2 \cos (2 \theta )+e^2+4\),
\n
while
\m
\left\langle\dot{L}_{EM}^1\right\rangle=\left\langle\dot{L}_{EM}^2\right\rangle=0.
\n

Similarly to electromagnetic radiation, for gravitational radiation, the averaged energy loss rate is given by
\m
\label{B3}
&&\left\langle\frac{dE_{GW}}{dt}\right\rangle=\frac{(-C)^3 \csc ^4(\theta )}{240 a^5 \left(1-e^2\right)^{7/2}  \mu }
\notag \\
&\times& \Bigl( 2 \left(e^2+1\right) \left(15 e^2+308\right) e^2 (\cos (4 \theta ))
\notag \\
&+&\left(-15 e^6+26 e^4+1976 e^2+720\right) (\cos (2 \theta ))
\notag \\
&-&3 \left(15 e^6+404 e^4+1104 e^2+272\right) \Bigr),
\n
and the averaged angular momentum loss rate by
\m
\label{B4}
&&\left\langle\dot{L}_{GW}^3\right\rangle=\frac{(-C)^{5/2 } \csc ^3(\theta )}{40 a^{7/2} \sqrt{\mu }(1-e^2)^2 }
\notag \\
&\times& \Bigl( \left(7 e^2+48\right) e^2 (\cos (4 \theta ))-2 \left(5 e^4+92 e^2+68\right)
\notag \\
&+&\left(-3 e^4+88 e^2+120\right) \cos (2 \theta ) \Bigr),
\n
while
\m
\left\langle\dot{L}_{GW}^1\right\rangle=\left\langle\dot{L}_{GW}^2\right\rangle=0.
\n

Noting that if $\sin \theta =l/n$ and using
\m
\cos (2 \theta )=1-\frac{2 l^2}{n^2},  \quad
\cos (4 \theta )=\frac{8 l^4}{n^4}-\frac{8 l^2}{n^2}+1,
\n
we can show that Eqs.~\eqref{A1}, \eqref{A2}, \eqref{A3} and \eqref{A4} are consistent with Eqs.~\eqref{B1}, \eqref{B2}, \eqref{B3} and \eqref{B4}. Here, we have shown that regardless of rational or irrational $\sin \theta$, the emission rates of energy and angular momentum due to gravitational and electromagnetic radiation have the same form as expected. Intuitively we note that the emissions of energy and angular momentum can continuously change $\sin \theta$ from irrational values to rational values or vice versa. So, the emission rates of energy and angular momentum due to gravitational and electromagnetic radiation should have the same dependence on $\theta$ no matter how $\sin \theta$ is rational or irrational. Now we show that Eqs.~\eqref{B1}, \eqref{B2}, \eqref{B3} and \eqref{B4} are also valid for the case $e=0$:
\e
\a
\left\langle\frac{dE_{EM}}{dt}\right\rangle=-\frac{2 ((\Delta \sigma_q)^{2}+(\Delta \sigma_g)^{2}) (-C) \left(-a C \mu +D^2\right)}{3 a^5 \mu },
\b
\q

\e
\a
\left\langle\frac{dL_{EM}}{dt}\right\rangle=-\frac{2 ((\Delta \sigma_q)^{2}+(\Delta \sigma_g)^{2}) (-C) \sqrt{-aC \mu +D^2}}{3 a^3},
\b
\q

\e
\a
\left\langle\frac{dE_{GW}}{dt}\right\rangle=-\frac{2 (-C) \left(-a C  \mu +D^2\right) \left(-16 a C \mu +D^2\right)}{5 a^7 \mu ^3},
\b
\q

\e
\a
\left\langle\frac{dL_{GW}}{dt}\right\rangle=-\frac{2 (-C) \sqrt{-a C \mu +D^2} \left(-16 a C \mu +D^2\right)}{5 a^5 \mu ^2}.
\b
\q
These are consistent with \cite{Liu:2020vsy} by using $\sin \theta=\frac{\sqrt{-a C \mu}}{\sqrt{-a C \mu +D^2}}$.

In \cite{Liu:2020vsy}, the static orbits (without radiation) are divided into three categories: (1) $e=0$; (2) $e \neq 0$, $\sin \theta$ is rational; (3) $e \neq 0$, $\sin \theta$ is irrational. In this section, we have shown that the emission rates of energy and angular momentum have the same dependence on $\theta$ for all cases. Finally, we conclude that no matter how $\sin \theta$ is rational or irrational and $e=0$ or not, the total emission rates of energy and angular momentum due to gravitational and electromagnetic radiation are
\m
\label{dEtotal}
\left\langle\frac{dE}{dt}\right\rangle=\left\langle\frac{dE_{EM}}{dt}\right\rangle+\left\langle\frac{dE_{GW}}{dt}\right\rangle,
\n
and
\m
\label{dLtotal}
\left\langle\frac{dL}{dt}\right\rangle=\left\langle\frac{dL_{EM}}{dt}\right\rangle+\left\langle\frac{dL_{GW}}{dt}\right\rangle,
\n
where $\left\langle\frac{dE_{EM}}{dt}\right\rangle$, $\left\langle\frac{dL_{EM}}{dt}\right\rangle$, $\left\langle\frac{dE_{GW}}{dt}\right\rangle$ and $\left\langle\frac{dL_{GW}}{dt}\right\rangle$ are given by Eqs.~\eqref{B1}, \eqref{B2}, \eqref{B3} and \eqref{B4}, respectively.

\section{Evolutions of orbits}
\label{Evolution}
Now that we have the emissions of energy and angular momentum due to gravitational radiation and electromagnetic radiation, we can be able to calculate the evolution of the orbit through two Keplerian parameters $a$, $e$ and another parameter $\theta$ due to the presence of a magnetic charge.

\begin{figure}[htpb]
    \includegraphics[width=0.48\textwidth]{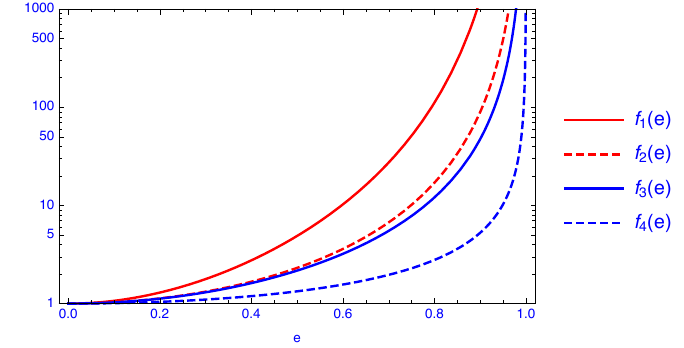}
\caption[]{The plots of $f_1(e)$, $f_2(e)$, $f_3(e)$ and $f_4(e)$ as functions of $e$ by choosing $m_1=m_2=m$, $q_1=q_2=0.2m$, $g_1=-g_2=0.1m$ and $a=10^4 m$.}
\label{fig:E}
\end{figure}

Though we have the emissions of energy ~\eqref{dEtotal} and angular momentum~\eqref{dLtotal}, we have three parameters $a, e,$ and $\theta$. According to Subsection \ref{Rational} and \ref{Irrational}, when we consider gravitational and electromagnetic radiation, we find only $\dot{L}^3 \neq 0$, which implies that the direction of $\bm{L}$ does not change while the magnitude of $\bm{L}$ decreases. From Eq.~\eqref{tilde} and $\tan(\theta)=\tilde{L}/|D|$, we have the relation of $a$, $e$ and $\theta$:
\m
\label{Relation}
\tan^2 (\theta)=\frac{\mu(-C)a(1-e^2)}{D^2}.
\n
Using the relation of $a$, $e$ and $\theta$, we may rewrite Eqs.~\eqref{dEtotal} and \eqref{dLtotal} as functions of $a$ and $e$ as
\begin{widetext}
\e
\a
\left\langle\frac{dE}{dt}\right\rangle &=& \frac{((\Delta \sigma_q)^{2}+(\Delta \sigma_g)^{2}) (-C) \left(4 a \left(e^4+e^2-2\right) (-C) \mu -D^2 \left(3 e^4+24 e^2+8\right)\right)}{12 a^5 \left(1-e^2\right)^{7/2} \mu }
 \\
&&-\frac{(-C)^{5/2} \left(8 a^2 h_1 C^2 \mu ^2- 3 a D^2 h_2 (-C) \mu +3 D^4 h_3\right)}{120 a^{11/2} \left(1-e^2\right)^4 \mu ^{3/2} \left(D^2+a \left(1-e^2\right) (-C) \mu \right)^{3/2}},
\b
\q
\e
\a
\left\langle\frac{dL}{dt}\right\rangle &=& -\frac{((\Delta \sigma_q)^{2}+(\Delta \sigma_g)^{2}) (-C) \left(D^2 \left(e^2+2\right)+2 a \left(1-e^2\right) (-C) \mu \right)}{3 a^3 \left(1-e^2\right)^{3/2} \sqrt{ \left(D^2+a \left(1-e^2\right) (-C) \mu \right)}} \\
&&-\frac{(-C) \left(16 a^2  h_4 C^2 \mu ^2 - a D^2 h_5 (-C) \mu +D^4h_6 \right)}{20 a^5 \left(1-e^2\right)^{7/2} \mu ^2 \sqrt{ \left(a \left(1-e^2\right) (-C) \mu +D^2\right)}},
\b
\q
\end{widetext}
where the first term is the energy (angular momentum) loss rate due to electromagnetic radiation and the second term is the energy (angular momentum) loss rate due to gravitational radiation. Here, we have used the short-hand notations
\begin{eqnarray}
h_1 &=& \left(1-e^2\right)^2 (37 e^4+292 e^2+96), \nonumber\\
h_2 &=& 45 e^8+1005 e^6+670 e^4-1448 e^2-272,
\nonumber\\
h_3 &=& 5 e^6+90 e^4+120 e^2+16
\end{eqnarray}
and
\begin{eqnarray}
h_4 &=& \left(1-e^2\right)^2 (7 e^2+8), \nonumber\\
h_5 &=& 31 e^6+297 e^4-192 e^2-136, \nonumber\\
h_6 &=& 3 (e^2+8) e^2+8.
\end{eqnarray}
In principle, we can also rewrite Eqs.~\eqref{dEtotal} and \eqref{dLtotal} as functions of $a$ and $\theta$ or functions of $e$ and $\theta$.
%For example, we rewrite Eqs.\eqref{dEtotal} and \eqref{dLtotal} as functions of $a$ and $\theta$ in Appendix.
When we rewrite Eqs.~\eqref{dEtotal} and \eqref{dLtotal} as functions of Keplerian parameters $a$ and $e$, it is much more easily to come back to the results of \cite{Liu:2020cds,Peters:1964zz}.
To compare the differences of the energy (angular momentum) loss rate between circular orbits and elliptical orbits, we define
\m
f_1(e) = \left\langle\frac{dE_{GW}}{dt}\right\rangle/\left\langle\frac{dE_{GW}}{dt}\right\rangle|_{e=0},
\n
\m
f_2(e) = \left\langle\frac{dE_{EM}}{dt}\right\rangle/\left\langle\frac{dE_{EM}}{dt}\right\rangle|_{e=0},
\n
and
\m
f_3(e) = \left\langle\frac{dL_{GW}}{dt}\right\rangle/\left\langle\frac{dL_{GW}}{dt}\right\rangle|_{e=0},
\n
\m
f_4(e) = \left\langle\frac{dL_{EM}}{dt}\right\rangle/\left\langle\frac{dL_{EM}}{dt}\right\rangle|_{e=0}.
\n
In Fig.~\ref{fig:E}, we plot $f_1(e)$, $f_2(e)$, $f_3(e)$ and $f_4(e)$ as functions of $e$ by choosing $m_1=m_2=m$, $q_1=q_2=0.2m$, $g_1=-g_2=0.1m$ and $a=10^4 m$. From Fig.~\ref{fig:E}, {we find that the energy and angular momentum loss rates increase quite fast as the eccentricity increases. Thus, highly elliptical orbits lose the energy and angular momentum more rapidly than the less elliptical orbits.}

Now, we find the evolution of $a$ and $e$. From the chain rule for differentiation, we have
\m
\frac{dE}{da}\frac{da}{dt}=\left\langle\frac{dE}{dt}\right\rangle,
\n
\m
\frac{dL}{da}\frac{da}{dt}+\frac{dL}{de}\frac{de}{dt}=\left\langle\frac{dL}{dt}\right\rangle.
\n
For simplicity, we divide the rates of the semimajor axis and eccentricity into two parts: the first part is the loss rates due to electromagnetic radiation and the second part is the loss rates due to gravitational radiation. In other words,
\m
\label{dadt}
\frac{da}{dt}=\frac{da_{EM}}{dt}+\frac{da_{GW}}{dt},
\n
\m
\label{dedt}
\frac{de}{dt}=\frac{de_{EM}}{dt}+\frac{de_{GW}}{dt},
\n
where $\frac{da_{EM}}{dt}, \frac{da_{GW}}{dt}, \frac{de_{EM}}{dt} $ and $\frac{de_{GW}}{dt}$ satisfy, respectively,
\m
\frac{dE}{da}\frac{da_{EM}}{dt}=\left\langle\frac{dE_{EM}}{dt}\right\rangle,
\n
\m
\frac{dL}{da}\frac{da_{EM}}{dt}+\frac{dL}{de}\frac{de_{EM}}{dt}=\left\langle\frac{dL_{EM}}{dt}\right\rangle,
\n
and
\m
\frac{dE}{da}\frac{da_{GW}}{dt}=\left\langle\frac{dE_{GW}}{dt}\right\rangle,
\n
\m
\frac{dL}{da}\frac{da_{GW}}{dt}+\frac{dL}{de}\frac{de_{GW}}{dt}=\left\langle\frac{dL_{GW}}{dt}\right\rangle,
\n
Using
\m
E=\frac{C}{2a},
\n
and
\m
L= \sqrt{{a \left(1-e^2\right) (-C) \mu }+D^2},
\n
we finally obtain
\begin{widetext}
{%\footnotesize
%\tiny
\e
\label{daGW}
\a
\frac{da_{GW}}{dt}=\frac{(-C)^{3/2} \left(-8 a^2 \left(e^2-1\right)^2 h_1 C^2 \mu ^2+3 a D^2 h_2 (-C) \mu -3 D^4 h_3\right)}{60 a^{7/2} \left(e^2-1\right)^4 \left(\mu  \left(D^2-a \left(e^2-1\right) (-C) \mu \right)\right)^{3/2}},
\b
\q
\e
\label{deGW}
\begin{aligned}
\frac{de_{GW}}{dt}=-\frac{e \Bigl(8 a^2 h_7 C^2 \mu ^2 - 3 a D^2 h_8 (-C) \mu +33 D^4 h_9 \Bigr)}{120 a^6 \left(1-e^2\right)^{9/2} \mu ^3},
\end{aligned}
\q
\e
\label{daEM}
\a
\frac{da_{EM}}{dt}=-\frac{(\Delta \sigma_q)^{2}+(\Delta \sigma_g)^{2}) \Bigl(D^2 \left(3 e^4+24 e^2+8\right)-4 a \left(e^4+e^2-2\right) (-C) \mu \Bigr)}{6 a^3 \left(1-e^2\right)^{7/2} \mu },
\b
\q
\e
\label{deEM}
\a
\frac{de_{EM}}{dt}=-\frac{((\Delta \sigma_q)^{2}+(\Delta \sigma_g)^{2}) e \left(7 D^2 \left(e^2+4\right)+12 a \left(1-e^2\right) (-C) \mu \right)}{12 a^4 \left(1-e^2\right)^{5/2} \mu },
\b
\q
}
\end{widetext}
where
\begin{eqnarray}
h_7 &=& \left(1-e^2\right)^2 \left(121 e^2+304\right), \nonumber\\
h_8 &=& 107 e^6+1537 e^4-308 e^2-1336, \nonumber\\
h_9 &=& e^4+12 e^2+8.
\end{eqnarray}

Using Eqs.~\eqref{dadt}, \eqref{dedt}, \eqref{daGW}, \eqref{deGW}, \eqref{daEM} and \eqref{deEM}, we can obtain $\frac{da}{dt}$ and $\frac{de}{dt}$ as functions of $a$ and $e$ which means we find the evolutions of orbits. Notice that $0\leq e <1$ and $C<0$, for arbitrary $q_1, q_2, g_1, g_2, m_1$ and $m_2$, we always have $\frac{da_{GW}}{dt}<0$, $\frac{de_{GW}}{dt} \leq 0 $, $\frac{da_{EM}}{dt}<0$ and $\frac{de_{EM}}{dt} \leq 0$ which imply $\frac{da}{dt}<0$ and $\frac{de}{dt} \leq 0$. If $e=0$, we have
\e
\a
\frac{da_{GW}}{dt}=-\frac{4 \left(a (-C) \mu +D^2\right) \left(16 a (-C) \mu +D^2\right)}{5 a^5 \mu ^3},
\b
\q
\e
\a
\frac{da_{EM}}{dt}=-\frac{4 ((\Delta \sigma_q)^{2}+(\Delta \sigma_g)^{2}) \left(a (-C) \mu +D^2\right)}{3 a^3 \mu },
\b
\q
\e
\a
\frac{de_{GW}}{dt}=\frac{de_{EM}}{dt}=0,
\b
\q
which are consistent with \cite{Liu:2020vsy}. Therefore a circular orbit remains circular. For $e>0$, we have $\frac{de}{dt}=\frac{de_{GW}}{dt}+\frac{de_{EM}}{dt}<0$ instead, and therefore an elliptical orbit becomes quasi-circular because of electromagnetic and gravitational radiations. Here, we notice that
\m
\frac{da_{GW}}{dt}/\frac{da_{GW}}{dt}|_{e=0}= \left\langle\frac{dE_{GW}}{dt}\right\rangle/\left\langle\frac{dE_{GW}}{dt}\right\rangle|_{e=0}=f_1(e),
\notag
\\
\n
\m
\frac{da_{EM}}{dt}/\frac{da_{EM}}{dt}|_{e=0} = \left\langle\frac{dL_{GW}}{dt}\right\rangle/\left\langle\frac{dL_{GW}}{dt}\right\rangle|_{e=0}=f_3(e).
\notag
\\
\n

 In this subsection, we obtain the evolutions of orbits in three-dimensional trajectories. Next, we will show our results are also valid for orbits confined in $x$-$y$ plane, in other words, $D=0$, which corresponds to purely electric or magnetic charges or $q_2/q_1= g_2/g_1$  of balancing out the velocity-dependent Lorentz forces. When $D=0$, we find
\e
\a
\frac{da_{GW}}{dt}=-\frac{2 \left(37 e^4+292 e^2+96\right) C^2}{15 a^3 \left(1-e^2\right)^{7/2} \mu },
\b
\q
\e
\a
\frac{de_{GW}}{dt}=-\frac{e \left(121 e^2+304\right) C^2}{15 a^4 \left(1-e^2\right)^{5/2} \mu },
\b
\q
\e
\a
\frac{da_{EM}}{dt}=-\frac{2 ((\Delta \sigma_q)^{2}+(\Delta \sigma_g)^{2})\left(e^2+2\right) (-C)}{3 a^2 \left(1-e^2\right)^{5/2}},
\b
\q
\e
\a
\frac{de_{EM}}{dt}=-\frac{((\Delta \sigma_q)^{2}+(\Delta \sigma_g)^{2}) e (-C)}{a^3 \left(1-e^2\right)^{3/2}},
\b
\q
which are consistent with \cite{Liu:2020cds,Christiansen:2020pnv}. In particular, for Schwarzschild black holes, $g_1=g_2=q_1=q_2=0$, which imply $\Delta \sigma_q=\Delta \sigma_g=D=0$ and $C=-\mu M$. Then we get
\m
\frac{da_{GW}}{dt}=-\frac{2 \left(37 e^4+292 e^2+96\right) \mu  M^2}{15 a^3 \left(1-e^2\right)^{7/2}},
\n
\m
\frac{de_{GW}}{dt}=-\frac{e \left(121 e^2+304\right) \mu  M^2}{15 a^4 \left(1-e^2\right)^{5/2}},
\n
\m
\frac{da_{EM}}{dt}=\frac{de_{EM}}{dt}=0,
\n
which are consistent with \cite{Peters:1963ux,Peters:1964zz}.
According to Section \ref{Evolution} and Appendix, we can conclude a few features of the evolution of orbits.
\begin{enumerate}
\item For arbitrary cases, the semimajor axis $a$ always decreases with time due to the energy loss in gravitational and electromagnetic waves.
\item A circular orbit remains circular while an elliptical orbit becomes quasi-circular due to the loss of energy and angular momentum. In other words, the effect of the back-reaction of gravitational and electromagnetic waves is to circularize the orbit.
\item
When $D=0$, the conic angle $\theta$ keeps the value $\theta=\pi/2$ which means the orbit is confined in $x$-$y$ plane.
When $D\neq0$, with the semimajor axis $a$ shrinking, the conic angle $\theta$ decreases. When the semimajor axis $a$ shrinks to  nearly zero, the conic angle $\theta$ also decreases to nearly zero.
\end{enumerate}

\section{ Waveform for dyonic binary black hole inspirals}
\label{Waveformcompare}

{ In \Sec{Emission} and \Sec{Evolution}, we have calculated the total emission rate of energy and angular momentum due to the gravitational and electromagnetic radiations from binary black holes with electric and magnetic charges and obtained the evolutions of the orbits. In this section, we will derive the waveforms for dyonic binary black hole inspirals.}

Following \cite{Maggiore:1900zz}, for the waveform emitted into an arbitrary direction $\bm{\vec{n}}$, which we can set it as
\m
\vec{\bm{n}}=(\sin \theta_1 \sin \phi_1, \sin \theta_1 \cos \phi_1, \cos\theta_1),
\n

we have
\begin{widetext}
\m
h_{+}&=&\frac{1}{d}(\ddot{M}_{11}(\cos^2\phi_1-\sin^2\phi_1\cos^2\theta_1)+\ddot{M}_{22}(\sin^2\phi_1-\cos^2\phi_1\cos^2\theta_1)-\ddot{M}_{33} \sin^2\theta_1
\notag\\
&-& \ddot{M}_{12} \sin{2\phi_1}(1+\cos^2\theta_1)+\ddot{M}_{13}\sin\phi_1\sin2\theta_1+\ddot{M}_{23}\cos\phi_1\sin2\theta_1),
\n
\m
h_\times=\frac{1}{d}((\ddot{M}_{11}-\ddot{M}_{22})\sin2\phi_1\cos\theta_1+2\ddot{M}_{12}\cos2\phi_1\cos\theta_1-2\ddot{M}_{13}\cos\phi_1\sin\theta_1+2\ddot{M}_{23}\sin\phi_1\sin\theta_1),
\n
where $d$ is the distance from the dyonic black hole to the earth. According to \eqref{Mij}, we obtain
\begin{equation}
\label{wave1}
\begin{array}{l}
h_{+}=\frac{-1}{2 a\left(1-e^{2}\right) d}\left(-4 (-C)\left(\sin ^{2}(\phi_1)-\cos ^{2}(\theta_1) \cos ^{2}(\phi_1)\right)\left(e^{2} \sin (\theta) \sin (\phi) \cos (\phi) \sin (2 \phi \sin (\theta))+e \sin (\theta) \sin (2 \phi) \sin (\phi \sin (\theta))\right.\right. \\
+\left.\cos ^{2}(\phi)(e \cos (\phi \sin (\theta))+1)^{2}-\frac{1}{2} \sin ^{2}(\phi)(e(\cos (2 \theta)(e+\cos (\phi \sin (\theta)))+e \cos (2 \phi \sin (\theta))+3 \cos (\phi \sin (\theta)))+2)\right) \\
+\frac{1}{2} (-C)(\cos (2 \theta_1)+3) \sin (2 \phi_1)(4 e \sin (\theta) \cos (2 \phi) \sin (\phi \sin (\theta))(e \cos (\phi \sin (\theta))+1) \\
-\left.\sin (2 \phi)\left(e^{2}+e(\cos (2 \theta)(e+\cos (\phi \sin (\theta)))+2 e \cos (2 \phi \sin (\theta))+7 \cos (\phi \sin (\theta)))+4\right)\right) \\
+(-C) \sin (2 \theta_1) \sin (\phi_1)\left(\cot (\theta) \cos (\phi)\left(-e^{2}+e(2 \cos (2 \theta)(e+\cos (\phi \sin (\theta)))+e \cos (2 \phi \sin (\theta))+2 \cos (\phi \sin (\theta)))+2\right)\right. \\
+4 e \cos (\theta) \sin (\phi) \sin (\phi \sin (\theta))(e \cos (\phi \sin (\theta))+1))\\
+(-C) \sin (2 \theta_1) \cos (\phi_1)\left(\cot (\theta) \sin (\phi)\left(-e^{2}+e(2 \cos (2 \theta)(e+\cos (\phi \sin (\theta)))+e \cos (2 \phi \sin (\theta))+2 \cos (\phi \sin (\theta)))+2\right)\right. \\
-4 e \cos (\theta) \cos (\phi) \sin (\phi \sin (\theta))(e \cos (\phi \sin (\theta))+1))+2\left(\cos ^{2}(\phi_1)-\cos ^{2}(\theta_1) \sin ^{2}(\phi_1)\right) \\
\times \left((-C) \cos ^{2}(\phi)(e(\cos (2 \theta)(e+\cos (\phi \sin (\theta)))+e \cos (2 \phi \sin (\theta))+3 \cos (\phi \sin (\theta)))+2)-2 (-C)(e \cos (\phi \sin (\theta))+1)\right. \\
\times \left(\sin ^{2}(\phi)(e \cos (\phi \sin (\theta))+1)-e \sin (\theta) \sin (2 \phi) \sin (\phi \sin (\theta))\right)+4 e (-C) \cos ^{2}(\theta) \sin ^{2}(\theta_1)(e+\cos (\phi \sin (\theta))),
\end{array}
\end{equation}

\begin{equation}
\label{wave2}
\begin{array}{l}
h_{\times}=\frac{C}{2 a\left(1-e^{2}\right) d} \left(2 \cos (\theta_1) \cos ^{2}(\phi) \sin (2 \phi_1)\left(e^{2}+e(\cos (2 \theta)(e+\cos (\phi \sin (\theta)))+2 e \cos (2 \phi \sin (\theta))+7 \cos (\phi \sin (\theta)))+4\right)\right. \\
+\cos (\theta_1)\left(\sin (\phi)(3 \cos (\phi+2 \phi_1)+\cos (\phi-2 \phi_1))\left(e^{2}+e(\cos (2 \theta)(e+\cos (\phi \sin (\theta)))+2 e \cos (2 \phi \sin (\theta))+7 \cos (\phi \sin (\theta)))+4\right)\right. \\
-8 e \sin (\theta) \sin (\phi \sin (\theta)) \cos (2(\phi+\phi_1))(e \cos (\phi \sin (\theta))+1)) \\
-2 \cot (\theta) \sin (\theta_1) \cos (\phi) \cos (\phi_1)\left(-e^{2}+e(2 \cos (2 \theta)(e+\cos (\phi \sin (\theta)))+e \cos (2 \phi \sin (\theta))+2 \cos (\phi \sin (\theta)))+2\right) \\
+2 \sin (\theta_1)\left(\cot (\theta) \sin (\phi) \sin (\phi_1)\left(-e^{2}+e(2 \cos (2 \theta)(e+\cos (\phi \sin (\theta)))+e \cos (2 \phi \sin (\theta))+2 \cos (\phi \sin (\theta)))+2\right)\right. \\
-4 e \cos (\theta) \sin (\phi \sin (\theta)) \sin (\phi+\phi_1)(e \cos (\phi \sin (\theta))+1))),
\end{array}
\end{equation}
where the evolutions of $a, e$ are determined by Eqs. \eqref{dadt}, \eqref{dedt}, \eqref{daGW}, \eqref{deGW}, \eqref{daEM} and \eqref{deEM}, and $\theta$ is a function of $a$ and $e$ which is given by \eqref{Relation}. Equations \eqref{wave1} and \eqref{wave2} are the waveforms for dyonic binary black hole inspirals. {In \Sec{Evolution}, we have shown that an elliptic orbit becomes quasi-circular due to the radiation reaction of the gravitational and electromagnetic radiations. Particularly, for circular obits ($e=0$), we can rewrite Eqs. \eqref{wave1} and \eqref{wave2} in simple forms,
\m
\label{h1e=0}
h_{+}=- \frac{\sqrt{-C}}{\sqrt{a^3 \mu} d} \Bigl[|D| \sin (2 \theta_1 ) \sin (\phi +\phi_1) +\sqrt{(-C) a \mu} (\cos (2 \theta_1)+3) \cos (2 (\phi +\phi_1)) \Bigr],
\n
\m
\label{h2e=0}
h_\times= 2 \frac{\sqrt{-C}}{\sqrt{a^3 \mu} d} \Bigl[ |D| \sin (\theta_1) \cos (\phi +\phi_1) -2 \sqrt{(-C) a \mu} \cos (\theta_1) \sin (2 (\phi +\phi_1)) \Bigr],
\n
}
where the evolution of $a$ is given by
\m
\frac{da}{dt}=-\frac{4 \left(a (-C) \mu +D^2\right) \left(5 a^2 ((\Delta \sigma_q)^{2}+(\Delta \sigma_g)^{2}) \mu ^2+48 a (-C) \mu +3 D^2\right)}{15 a^5 \mu ^3}.
\n
\end{widetext}

\begin{figure}[htpb]
    \includegraphics[width=0.48\textwidth]{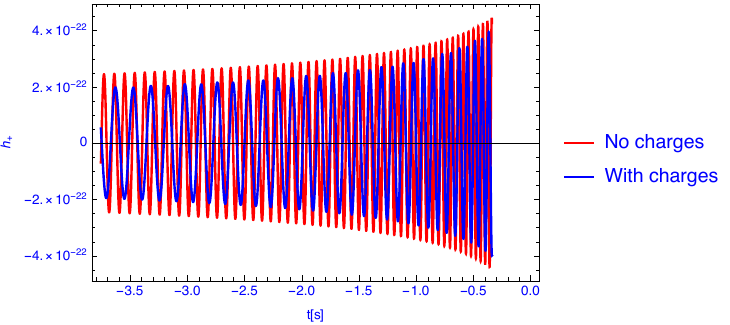}
    \includegraphics[width=0.48\textwidth]{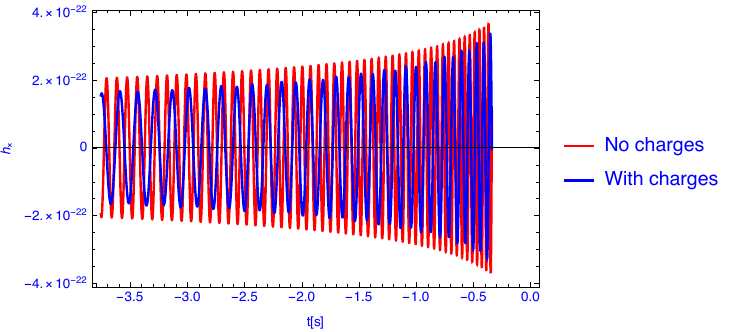}
    \caption[]{The plots of $h_{+}$ (top) and $h_{\times}$ (bottom) as functions of $t$ by choosing mass ratio $\frac{m_1}{m_2}=0.8$, chirp mass $\frac{(m_1m_2)^{3/5}}{(m_1+m_2)^{1/5}}=30\Msun$, $\theta_1=1\approx57.3^{\circ}$, $\phi_1=0$ and $d=450$Mpc. Red lines represent the case $q_1=q_2=g_1=g_2=0$ while blue lines represent the case $\frac{q_1}{m_1}=0.7, \frac{q_2}{m_2}=-0.1, \frac{g_1}{m_1}=-0.6, \frac{g_2}{m_2}=0.3$.}
    \label{fig: waveform}
\end{figure}

From \eqref{h1e=0} and \eqref{h2e=0}, we find when $D\neq0$, {the gravitational waveforms for dyonic binary black hole inspirals are superposed of } one waveform with frequency $\frac{\dot{\phi}}{2 \pi}$ and another waveform with frequency $\frac{\dot{\phi}}{\pi}$. It is a special feature of waveforms for dyonic binary black hole inspirals. For purely electric or magnetic or Schwarzschild black holes, the frequency of gravitational waveforms is $\frac{\dot{\phi}}{\pi}$ which is consistent with \cite{Maggiore:1900zz}. In Fig \ref{fig: waveform}, we explicitly show the difference between the waveform for dyonic binary black hole inspirals and that for Schwarzschild binary black hole inspirals. In this section, we have derived the waveforms for dyonic binary black hole inspirals. Those results can be applied to the black hole merger event to test the existence of electric and magnetic charges of black holes, { which will be explored in the next work}.

\section{conclusions and discussions}\label{Concl}
{In the universe, most binary black hole systems have a non-zero eccentricity. Binary black holes formed from encounters can even have large eccentricity. In fact, circular orbits are a very rare and special situation. Therefore, it is of great significance to explore the evolution and characteristics of the elliptical orbits of binary black holes with electric and magnetic charges.}
In this paper, we have investigated the equations of inspiral motion of dyonic binaries when their distance is much larger than their event horizons. By adopting the weak-field approximation and using a Newtonian method with radiation reactions included, we have calculated the total emission rates of energy and angular momentum due to gravitational and electromagnetic radiations from precessing elliptical orbits. It has been shown that the emission rates of energy and angular momentum due to gravitational and electromagnetic radiations have the same dependence on $\theta$ no matter how $\sin \theta$ is a rational number and closed orbit or an irrational number and chaotic orbit. Moreover, we have computed the evolution of orbits and found that a circular orbit remains circular while an elliptical orbit becomes more and more quasi-circular because of gravitational and electromagnetic radiations. Using the evolution of orbits, we have derived the waveform models for dyonic binary black hole inspirals. The results of this work can be used to investigate whether black holes have electric and magnetic charges or not.

{Within the framework of Newtonian orbits and the relativistic gravitational and electromagnetic radiations, we have only considered the leading orders of radiations and their effect on orbits: the post-Newtonian gravitational radiation and the synchrotron radiation due to electric and magnetic dipoles.} In other words, we have only considered 0-PN corrections. Even in the lowest-order Newtonian approximation, we have shown the amplitudes $h_+$ and $h_\times$ of the gravitational waves for dyonic binary black hole inspirals differ from those for Schwarzschild binary black hole inspirals. 

In the future, some aspects of the higher PN corrections for the orbits of dyonic black holes need to be discussed. In this paper, we have adopted the weak-field approximation. The main region of applicability of our results is the long inspirals that space-based GW detectors, such as LISA~\cite{Audley:2017drz} and Taiji ~\cite{Guo:2018npi}, will detect such gravitational waves.  When they nearly merge and the weak-field approximation breaks down, non-linear dynamics of charged binary black holes will play an important role and our results are not valid. A higher-order PN expansion or numerical-relativity simulations are needed, which will be studied in future works.
{ And the waveform model for dyonic binary black hole inspirals derived in this work is a part of the inspiral, merger, and ringdown (IMR) waveform model. The ringdown waveform for dyonic black holes is an interesting topic which we also leave for future works.}

\acknowledgments
{\O}C would like to thank Jose Beltr\' an Jim\' enez for discussions.  This work is supported in part by the National Key Research and Development Program of China Grant No.2020YFC2201501, in part by the National Natural Science Foundation of China Grants No.11690021,  No.11690022, No.11851302, No.11821505, No.11947302 and No.12075297, in part by the Strategic Priority Research Program of the Chinese Academy of Sciences Grant No. XDB23030100, No. XDA15020701 and by Key Research Program of Frontier Sciences, CAS.
The work of L.L. and S.P.K. also was supported in part by National Research Foundation of Korea (NRF) funded by the Ministry of Education (2019R1I1A3A01063183).

\appendix
\section*{Appendix: The evolution of $\theta$}
%\subsection{The evolution of $\theta$}
When $D=0$, which corresponds to purely electric or magnetic charges or $q_2/q_1= g_2/g_1$ of balancing out the velocity-dependent Lorentz forces, there is no angular-momentum-dependent and non-central force and $\bm{\tilde{L}}=\bm{L}$. According to $\dot{L}^1=\dot{L}^2=0$ while $\dot{L}^3 \neq 0$, the orbit is always confined in $x$-$y$ plane through the entire inspiral stage.
When $D \neq 0$, from
\m
L=|D|/\cos \theta,
\n
\m
\frac{dL}{d\theta}\frac{d\theta}{dt}=\left\langle\frac{dL}{dt}\right\rangle,
\n
we can obtain
\m
\frac{d\theta}{dt}=\left\langle\frac{dL}{dt}\right\rangle \frac{\cos (\theta ) \cot (\theta )}{|D|}.
\n
Because $\left\langle\frac{dL}{dt}\right\rangle<0$, we have $\frac{d\theta}{dt}<0$. According to \eqref{Relation}, when the semimajor axis $a$ shrinks to nearly zero, the conic angle $\theta$ also decreases to nearly zero.

%%%%%%%%%%%%%%%%%%%%%%%%%%%%%%%%%%%%%%%%
%%%%%%%%%%%%%%%%%%%%%%%%%%%%%%%%%%%%%%%%

%\bibliographystyle{apsrev}
\bibliography{merger}

%%%%%%%%%%%%%%%%%%%%%%%%%%%%%%%%%%%%%%%%
%%%%%%%%%%%%%%%%%%%%%%%%%%%%%%%%%%%%%%%%
\end{document}